\documentclass{article}
\usepackage{spconf,amsmath,graphicx}
\usepackage{color}
\usepackage{booktabs}
\usepackage{multirow}
\usepackage{subfigure}
\usepackage{bm}

\usepackage{enumitem}
\setlist{nosep, leftmargin=14pt}

\usepackage{mwe} % to get dummy images

\title{Mixing Data Augmentation with Preserving Foreground Regions in Medical Image Segmentation}

\name{Xiaoqing Liu \qquad Kenji Ono \qquad Ryoma Bise}

 \address{Graduate School and Faculty of Information Science and Electrical Engineering,\\ Kyushu University, Fukuoka, JAPAN \\
     }

\begin{document}

\maketitle

\begin{abstract}
The development of medical image segmentation using deep learning can significantly support doctors’ diagnoses. 
Deep learning needs large amounts of data for training, which also requires data augmentation to extend diversity for preventing overfitting.
However, the existing methods for data augmentation of medical image segmentation are mainly based on models which need to update parameters and cost extra computing resources. 
We proposed data augmentation methods designed to train a high-accuracy deep learning network for medical image segmentation. 
The proposed data augmentation approaches are called KeepMask and KeepMix, which can create medical images by better identifying the boundary of the organ with no more parameters. 
Our methods achieved better performance and obtained more precise boundaries for medical image segmentation on datasets.
The dice coefficient of our methods achieved 94.15\% (3.04\% higher than baseline) on CHAOS and 74.70\% (5.25\% higher than baseline) on MSD spleen with Unet. 

\end{abstract}
\section{Introduction}
\label{sec:intro}

Deep learning for medical image segmentation has a crucial role in obtaining quick and accurate results to aid diagnosis and prevent ``medical errors''. 
However, in medical image analysis, the amount of images in a dataset is often limited and may not be sufficient for training.
In this situation, improving the structure to be more complex and deeper often cause overfitting and learning of features that do not have commonality for small amounts of data.
In order to increase the diversity of data and make the distribution of datasets more continuous, data augmentation can realize improvements for enriching data and preventing overfitting.
For example, Eaton {\it et al.} \cite{eaton2018improving} applied Mixup\cite{zhang2017mixup} in medical images, and Xu {\it et al.} \cite{xu2020automatic} proposed an automatic augmentation for 3D medical image segmentation.
Although these proposed methods are effective, it is complex and costs extra computing resources while training models. 
These methods also need to update the parameters of the data augmentation methods depending on the datasets in addition to the model parameters of the segmentation network.
This will not only increase the memory burden but also extend the training time.
Simple data augmentation methods, which do not require searching for additional parameters, basically give perturbations ({\it e.g.}, GaussianBlur, GridDropout  \cite{buslaev2020albumentations}) to the entire region of an original image as augmented images. 
In medical image segmentation task, the image features of the boundary between the foreground and background regions are important, and thus such perturbations for foreground image may worse affect extracting discriminative features for segmentation.

In this paper, we propose two simple data augmentation methods for medical image segmentation, called KeepMask and KeepMix that can preserve the image features of the foreground regions (organ) but give perturbations for the background regions.
These methods are very simple but effective that do not require searching for additional parameters to fit different datasets and keep the vital features of the foreground regions.
The experimental results using three datasets; CHAOS \cite{kavur2021chaos,kavur2020comparison}, SLIVER07 \cite{heimann2009comparison}, and MSD spleen \cite{antonelli2021medical}, demonstrate the effectiveness of the proposed methods. 
Our approach achieved higher dice coefficients than baseline and prior arts and obtained more precise boundaries with few parameters and limited GPU memory.

\noindent
{\bf Related works:} In deep learning, data augmentation techniques are widely used in training models for image classification \cite{yun2019cutmix}, detection \cite{zoph2020learning}, segmentation \cite{ghiasi2021simple}, etc.
Gong et al. proposed Keep Augment \cite{gong2021keepaugment} that preserves the important region by a saliency map to avoid introducing noise by data augmentation. This method was designed for natural image classification. For medical image segmentation, if the images are generated only by the saliency map proposed by KeepAugment, the mask boundaries will be discontinuous, resulting in a low-performance model.
In the MedMix \cite{tian2021self}, they use strong data augmentation for contrastive learning for lesion detection by using cutting and pasting.
However, this approach is difficult to apply to medical image segmentation because the detection only needs to label the location of target while the segmentation needs to generate specific edges. 
We not only have adjusted the range of options and the tasks to be applied but also proposed a new mixing augmentation method applicable to medical image segmentation. 
\section{Method}
\label{sec:pagestyle}

Our methods preserve the vital features of the organ on medical images while transforming the background part on images. These methods generate images directly from the masks of the organ to fit different datasets and preserve the whole organ boundaries to allow the model to predict the boundaries of the organs more accurately.
We propose two approaches that follow the above idea.
We perform KeepMask on two basic data augmentation methods (GridDropout and GuassianBlur) from albumentations library \cite{buslaev2020albumentations} for KeepMask. Meanwhile, we also propose a simpler and more straightforward mixing approach called KeepMix instead of complex automatic data augmentation for easy application.

\begin{figure}
\centering
\includegraphics[width=8.5cm]{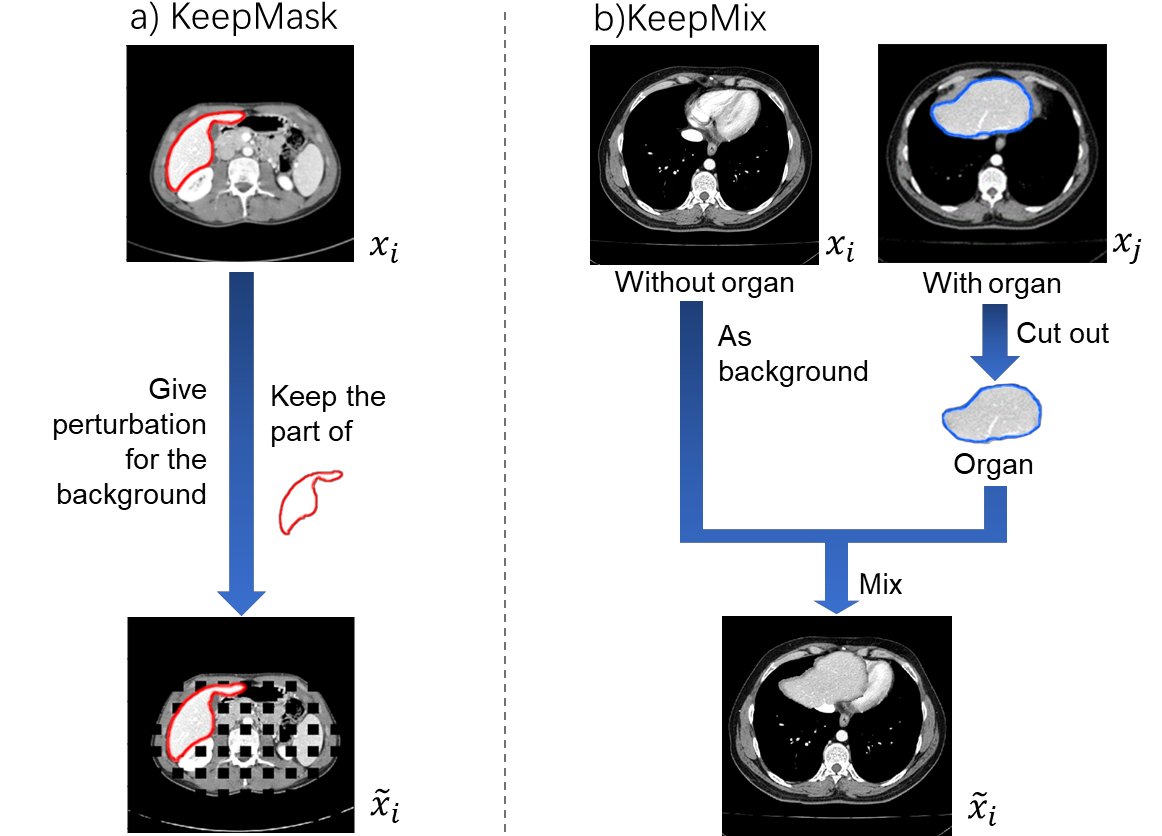} 
\caption{(a)KeepMask, and (b) KeepMix}
\label{fig:method}
\end{figure}

\subsection{KeepMask}
To focus on the features of the organ region and reduce the influence of other partial organs, KeepMask generates mixed augmented data that preserves the features of the foreground region.
Figure~\ref{fig:method}(a) shows the illustration of KeepMask. Given an input image $\bm{x}_i$ and the ground truth of segmentation $\bm{y}_i$, KeepMask adds a perturbations for the background regions while keeping the foreground regions.
The augmented image $\tilde{x}_i$ is defined as:
\begin{equation}
\tilde{\bm{x}}_i =\bm{y}_i\odot \bm{x}_i + (\bm{1}-\bm{y}_i)\odot \bm{x}_{i}', \hspace{3mm} x_{i}'=f_{DA}(x_i),
\end{equation}
where $\odot$ represents the multiplication of the corresponding positions of matrices with the same dimension, and $\bm{1}$ represents an image whose elements are 1 and size is the same size as $\bm{x}_i$.
Here, the ground truth of the segmentation mask for the augmented image $\tilde{\bm{y}}_i$ is the same with the original segmentation mask $\bm{y}_i$.
While any simple data augmentation method can be used the augmentation function for background $f_{DA}(\bm{i})$, we choose the GuassianBlur and GridDropout to imitate the difference in sharpness of different machine imaging.
Figure~\ref{fig:method}(a) shows an example of using GridDropout \cite{buslaev2020albumentations}.

\subsection{KeepMix}
%\noindent
%{\bf KeepMix:}
To generate the various background, KeepMix generates augmented images by mixing the pair of two images: one contains the foreground region (organ), and the other does not contain the foreground as shown in Figure~\ref{fig:method}(b).
Given the pair of supervised images ($\bm{x}_i,\bm{y}_i$), ($\bm{x}_j,\bm{y}_j$),
the augmented image $\tilde{\bm{x}}_i$ is defined as:
\begin{eqnarray}
\tilde{\bm{x}}_i &=&(\bm{1}-\bm{y}_i)\odot \bm{x}_i + \bm{y}_j\odot \bm{x}_j,
\end{eqnarray}
where, as same with KeepMask, $\tilde{\bm{y}}_i$ is the same with the original segmentation mask $\bm{y}_i$.

To generate the pairs $\bm{x}_i$, $\bm{x}_j$, we judge whether there is an organ in the image by the value of the mask.
If randomly selected image has an organ (foreground regions), we add the image as $\bm{x}_i$, and if the selected image does not have an organ, we add it as $\bm{x}_j$.
Moreover, we select the integration image among the images of the same patient as the original image.

In each iteration of training, it is considered better that samples in a batch contain both original and augmented images.
In both methods (KeepMask and KeepMix), we thus control whether or not to apply the proposed data augmentation method by giving the parameter $p$, which indicates the ratio of the original images in a batch.

The proposed method uses both KeepMask and KeepMix for data augmentation and trains any network using the augmented images.

\section{Experiments}
\label{sec:exp}

\subsection{Experimental Setup}
\textbf{Datasets.} We conducted the experiments on three datasets of CT scans, CHAOS \cite{kavur2021chaos,kavur2020comparison},  SLIVER07 \cite{heimann2009comparison}, and MSD spleen \cite{antonelli2021medical}. CHAOS \cite{kavur2021chaos,kavur2020comparison} contains CT scans of liver. We divided the train dataset into train set(8 patients), validation set(2 patients), and test set(10 patients) by patients. 
SLIVER07 \cite{heimann2009comparison} dataset contains 20 patients 3D CT scans of liver. For 2D segmentation, we sliced the 3D CT scans into 2D scans. We also divided the dataset into train set (8 patients), validation set (2 patients), and test set (10 patients). MSD spleen \cite{antonelli2021medical} dataset contains 40 patients 3D CT scans of spleen. We divided the dataset into train set (16 patients), validation set (4 patients), and test set (20 patients).

\begin{table}[!t] 
  \begin{center}
    \caption{Performance of compared methods on Unet and Unet++ under three datasets. The best performance in each block is highlighted in bold. $(...)$ indicates the improvements from the baseline.
    }\label{tab:result1}
    
    \resizebox{\columnwidth}{!}{
    \begin{tabular}{c|c|c|c|c} 
    \toprule
    Approach/(Dice\%)& Model&CHAOS& SLIVER07&MSD spleen\\%
    \midrule
    Baseline&\multirow{5}{*}{Unet}& 91.11&89.29 &69.45\\ 
    Basic DA&&93.11(+2.00)& 89.95(+0.66)&72.92(+3.47)\\
    Real-ESRGAN\cite{wang2021real}&&93.39(+2.28)&91.19(+1.90)&73.70(+4.25)\\
    RandomFog\cite{buslaev2020albumentations}&&92.70(+1.59)&90.83(+1.54)&71.67(+2.22)\\
    MaskDrop\cite{buslaev2020albumentations}&&93.54(+2.43)&90.93(+1.64)&71.22(+1.77)\\
    \midrule
    Ours&Unet& \textbf{94.15(+3.04)}&\textbf{91.87(+2.58)}&\textbf{74.70(+5.25)}\\
   \midrule
   \midrule
    Baseline&\multirow{5}{*}{Unet++}&92.09&90.83&70.23\\ 
    Basic DA&&94.14(+2.05)&91.54(+0.71)&74.17(+3.94)\\ 
    Real-ESRGAN\cite{wang2021real}&&93.65(+1.56)&91.69(+0.86)&76.92(+6.69)\\
    RandomFog\cite{buslaev2020albumentations}&&93.82(+1.73)&92.51(+1.68)&76.10(+5.87)\\
    MaskDrop\cite{buslaev2020albumentations}&&94.24(+2.15)&91.12(+0.29)&77.50(+7.27)\\
    \midrule
    Ours&Unet++&\textbf{94.72(+2.63)}&\textbf{92.90(+2.07)}&\textbf{77.64(+7.41)}\\
    \bottomrule
    \end{tabular} 
    }
    \end{center}
\end{table} 
\setlength{\tabcolsep}{1pt}

\noindent
\textbf{Segmentation Architectures.} Considering U-like network is most classical architecture in biomedical image segmentation, we use Unet \cite{ronneberger2015u} and Unet++ \cite{zhou2018unet++} with encoder of MobileNet v2 \cite{sandler2018mobilenetv2} as the segmentation architectures. We did not pretrain the network on any dataset.

\noindent
\textbf{Hyper-parameters.} For all the experiments, we set the hyper-parameters as follows. The batch size was 8 on Unet, and 4 on Unet++. The initial learning rate was set to 1e-4 and a step learning rate of 1/10. We trained the network for 50 epochs with early stopping. The hyper-parameter $p$ was selected using the validation data in each dataset.
\noindent
\textbf{KeepMask.} We applied KeepMask for improving two basic augmentation methods GuaussianBlur and GridDropout, which are called as \textbf{KeepGuassianBlur} and \textbf{KeepGridDropout}.

\noindent
\textbf{KeepMix.} In the KeepMix method, we performed on two modes. The first is to do KeepMix on the same patient with the sample, which is called \textbf{KeepMix(same)}. The second is to do KeepMix with a random sample from all patients, which is called \textbf{KeepMix(diff)}.

\noindent
\textbf{Performance metrics.} 
We used dice similarity coefficient (Dice) as an evaluation indicator. The dice coefficient is a measure of the overlap of the foreground regions in the estimated and ground truth segmentation images.

\subsection{Comparison of Augmentation methods}
Table~\ref{tab:result1} shows the results of the comparison of proposed methods and other related methods. All methods perform on Unet and Unet++ architectures to demonstrate the validity on different architectures. In \textbf{Baseline}, we do not use any data augmentation method. \textbf{Basic DA} means the classical data augmentation operations used in Unet are shift, rotation, and random elastic deformations. Meanwhile, we compared our methods with prior arts of \textbf{Real-ESRGAN} \cite{wang2021real}, \textbf{RandomFog} and \textbf{MaskDrop} \cite{buslaev2020albumentations}. In our methods, we combine KeepGridDropout, Keep GuassianBlur, and KeepMix(diff) together. Our methods perform better than all baseline, especially 5.25\% higher than baseline with Unet on MSD spleen and 7.41\% higher than baseline with Unet++ on MSD spleen. The above results strongly demonstrate the effectiveness of our methods.

\subsection{Comparison of single operation and baseline.}
    The effectiveness of KeepMask and KeepMix were verified separately on Unet as shown in table~\ref{tab:al}. 
    All of our methods improved the segmentation performance from Baseline.
    Our methods were more effective on the MSD spleen compared to the other dataset. This difference may be caused by the great differences among the images taken by amount of labeled data and Dice of results. When the model has already obtained high enough Dice, it is difficult to significantly improve the performance. Therefore, proposed data augmentation methods were more adapted for little labeled data and low Dice of Baseline.

   \setlength{\tabcolsep}{4pt}
    \begin{table}[!t] 
    \begin{center}
    \caption{Comparison of single operation and baseline.
    }\label{tab:al}
    
    \resizebox{\columnwidth}{!}{
    \begin{tabular}{c|c|c|c} 
    \toprule
    Dataset& CHAOS& SLIVER07&MSD spleen\\%
    \midrule
   Metrics & Dice(\%)& Dice(\%)&Dice(\%)\\
   \midrule
    Baseline&91.11&89.29&69.45\\ 
    \midrule
    KeepGuassianBlur& 93.13(+2.02)& 91.47(+2.18)&73.58(+4.13)\\
    KeepGridDropout& 92.88(+1.77)&90.65(+1.36)&74.10(+4.65)\\

    KeepMix(diff)& 93.61(+2.50)&90.74(+1.45)&72.92(+3.47)\\
    KeepMix(same)& 93.28(+2.17)& 91.21(+1.92)&73.61(+4.16)\\
    \midrule
    Ours&\textbf{94.15(+3.04)}&\textbf{91.87(+2.58)}&\textbf{74.70(+5.25)}\\
    \bottomrule
    \end{tabular} 
    }
    \end{center}
    %\vspace{-6mm}
    \end{table} 
    
\begin{figure}[t]
\begin{minipage}[b]{0.46\linewidth}
  \label{fig:keepgd}
  \centering
  \centerline{\includegraphics[width=4.6cm]{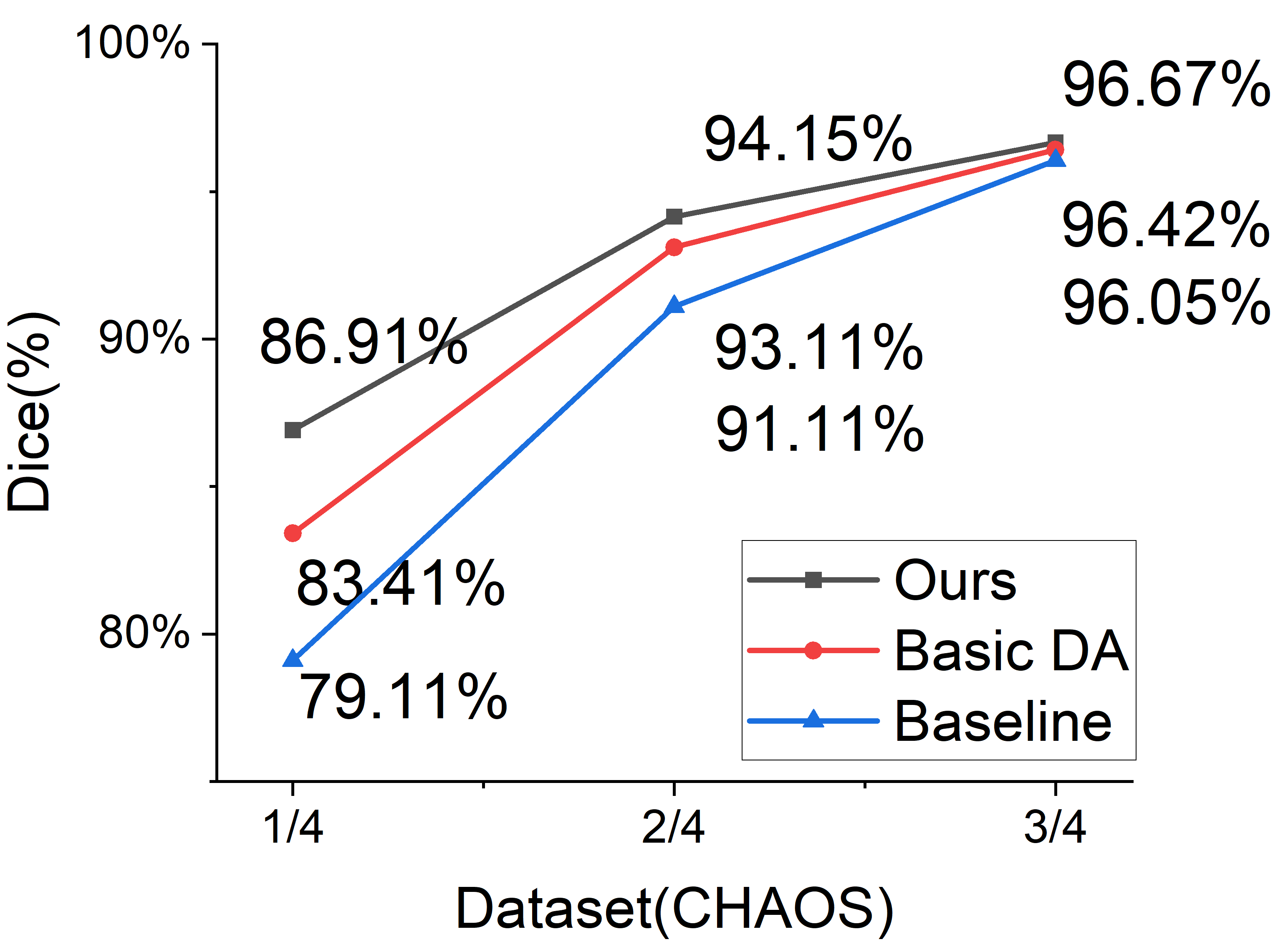}}
\end{minipage}
\hfill
\begin{minipage}[b]{0.46\linewidth}
  \label{fig:keepgb}
  \centering
  \centerline{\includegraphics[width=4.6cm]{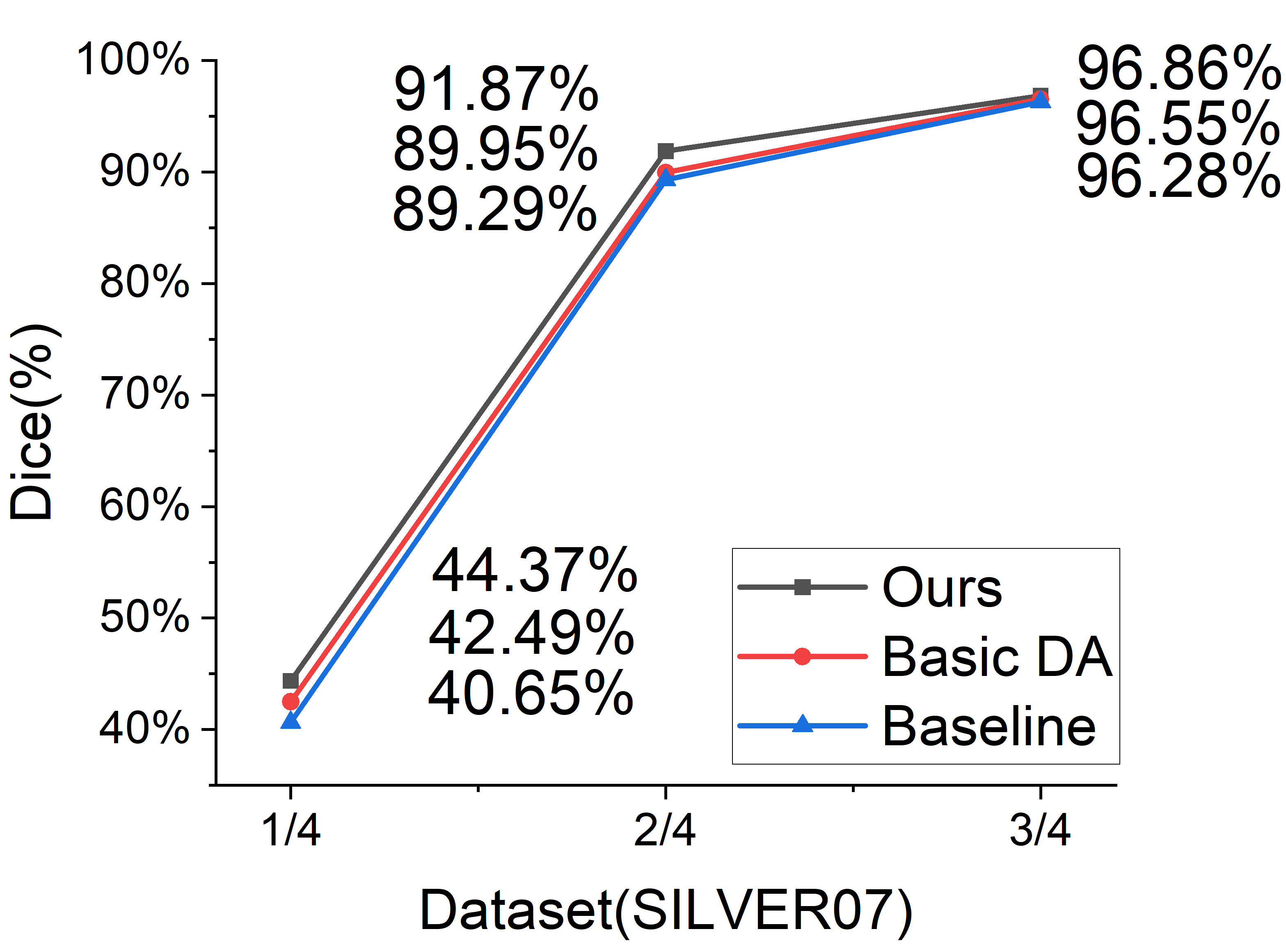}}
\end{minipage}
\caption{Comparison on different percentages of dataset CHAOS (Left) and SLIVER07 (Right). The vertical axis represents Dice. The horizontal axis represents the percentages(1/4, 1/2, 3/4) of dataset used for training.}
\label{fig:amount}
\end{figure}

 \subsection{Comparison of different amount of dataset}
    The improvement by data augmentation methods depends on the performance of Baseline and it depends on the number of the training data.
    We applied our method on 1/4, 1/2, 3/4 of datasets to demonstrate the validity of our methods on different percentages of labeled datasets, where we used Unet.
    Figure~\ref{fig:amount} shows that the smaller the percentages of the data set used for trainning, the more significant effect of our data augmentation method. Correspondingly, our approach always obtains the best performances.
    It shows that our methods are effective for medical image analysis, where it tends to be insufficient data.

\subsection{Robustness of KeepMask to the augmentation ratio.}
   We conduct the experiments to verify the effectiveness of KeepMask with different hyper-parameter $p$, which is the ratio of the augmented images in a batch, as shown in table~\ref{tab:KeepMask}.
   KeepMask improved the DSC with an average of 4\% from baseline on MSD spleen, and the DSCs of KeepMask were overall higher than those of Baseline and basic augmentation (GridDropout). It shows that the KeepMask is effective without depending on this hyper-parameter.

   \setlength{\tabcolsep}{1.4pt}
    \begin{table}[!t] 
    \begin{center}
    \caption{Effectiveness of KeepMask with different $p$, which is the ratio of applying augmentation.}\label{tab:KeepMask}
    \resizebox{\columnwidth}{!}{
    \begin{tabular}{c|c|c|c|c} 
    \toprule
   Approach/(Dice\%)& $p$ &CHAOS&SLIVER07&MSD spleen\\
   \midrule
    Baseline& -&91.11&89.29&69.45\\ 
    \midrule
    GridDropout& \multirow{2}{*}{0.1}&92.06(+0.95)&90.16(+0.87)&73.05(+3.60)\\ 
    KeepGridDropout& & \textbf{92.88(+1.77)}&\textbf{90.65(+1.36)}&\textbf{74.10(+4.65)}\\
    \midrule
    GridDropout& \multirow{2}{*}{0.2}&92.52(+1.41)&91.04(+1.75)&72.84(+3.39)\\ 
    KeepGridDropout&  & \textbf{92.58(+1.47)}&\textbf{91.66(+2.37)}&\textbf{73.88(+4.43)}\\
    \midrule
    GridDropout& \multirow{2}{*}{0.3}&92.64(+1.53)&88.76(-0.53)&67.79(-1.66)\\ 
    KeepGridDropout&  & \textbf{93.35(+2.24)}&\textbf{90.59(+1.30)}&\textbf{72.37(+2.92)}\\
    \bottomrule
    \end{tabular} 
    }
    \end{center}
    \vspace{-3mm}
    \end{table}
    \setlength{\tabcolsep}{1pt}

\begin{figure}[t]
\centering
\includegraphics[width=\linewidth]{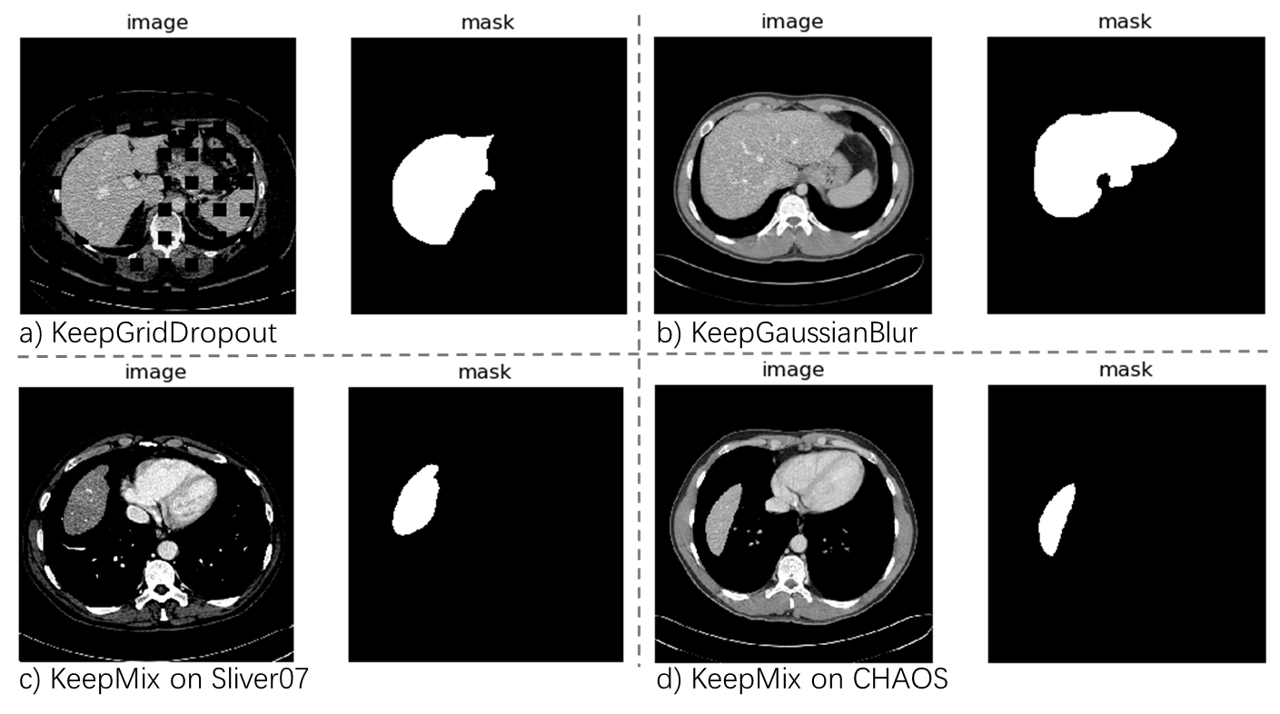} 
\caption{\textbf{Mixed Image:} (a) Keep part of mask and do GridDropout for background by KeepMask (b) Keep part of mask and do GuassianBlur for background by KeepMask (c) Mixed images and masks on SLIVER07 by KeepMix (d) Mixed images and masks on CHAOS by KeepMix}
\label{fig:miximage}
\end{figure}

\begin{figure}[ht]
\centering
\includegraphics[width=\linewidth]{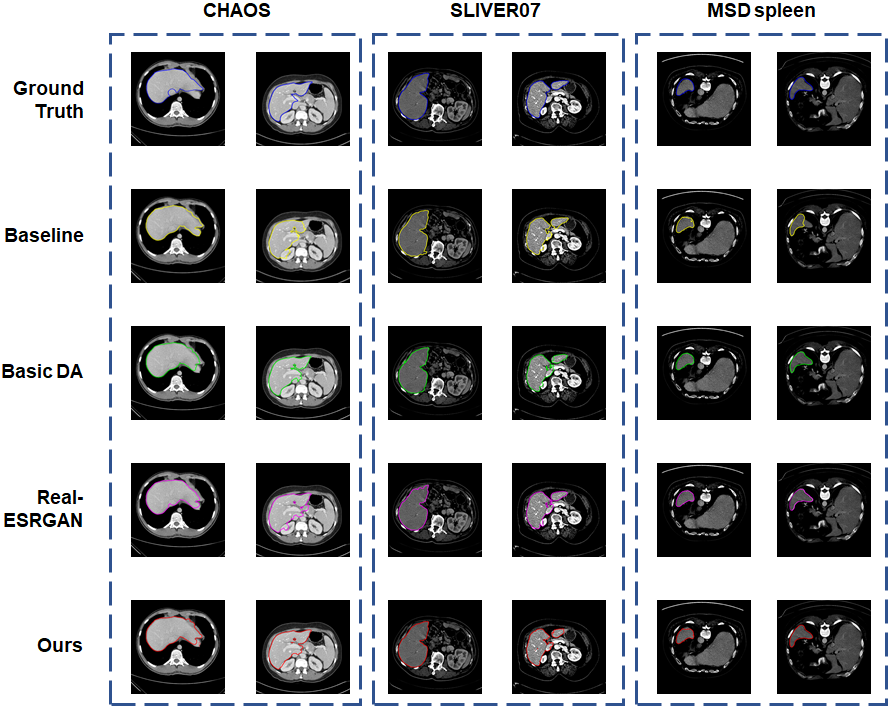}
\caption{
Visualizing boundaries of predicted masks by different methods. We randomly selected two images from each dataset to visualize the predicted boundaries.}
\label{fig:compareresults}
\end{figure}

\subsection{Visualization}
    Fig~\ref{fig:miximage} shows examples of augmented images and these masks of KeepMask ((a),(b)) and KeepMix ((c),(d)).
    In KeepMask, we can observe that the foreground regions (liver) clearly appeared in the augmented images while the perturbations (Gaussian blur and GridDropout) in the background regions. 
    In KeepMix, we can see that the image with the liver is generated on the image without the liver and it fits the distribution of the image.

    Fig.~\ref{fig:compareresults} shows examples of the segmentation results by each method.
    The first line is the ground truth of masks. The second line is the model without data augmentation. Basic DA and Real-ESRGAN are selected as prior arts for comparison. The last line is our method. Although the overall position of the segmentation is roughly the same, the segmentation boundary is more accurate with the proposed data augmentation method than without it.

\section{Conclusion}
\label{con}

 In this paper, we proposed simple and effective data augmentation methods for medical image segmentation, which increase the diversity of data to improve the performance of the model. The proposed KeepMask preserved the essential part of medical images and KeepMix mixes images with object and background from other images to generate new images with organ for data augmentation. 
 We performed the experiments to demonstrate the effectiveness of KeepMask and KeepMix on CHAOS, SLIVER07 and MSD spleen. Our method was better than the comparative methods, and the improvement by our method was large when the number of training data was few.
 It shows that our method is effective for medical image analysis, where data tends to be insufficient.

\noindent
\section{Compliance with ethical standards}
\label{sec:ethics}
This research study was conducted retrospectively using
    human subject data made available in open access by (Source
    information). Ethical approval was not required as confirmed by
    the license attached with the open access data.

\noindent
\section{Acknowledgments}
\par
\noindent
This work was supported by JSPS KAKENHI Grant Number JP20H04211, Japan. 
This work was supported by AMED under Grant Number JP20he2302002. 

\bibliographystyle{IEEEbib}
\bibliography{ISBI_latex}

\begin{thebibliography}{10}

\bibitem{eaton2018improving}
Zach Eaton-Rosen, Felix Bragman, Sebastien Ourselin, and M~Jorge Cardoso,
\newblock ``Improving data augmentation for medical image segmentation,''
\newblock 2018.

\bibitem{zhang2017mixup}
Hongyi Zhang, Moustapha Cisse, Yann~N Dauphin, and David Lopez-Paz,
\newblock ``mixup: Beyond empirical risk minimization,''
\newblock {\em arXiv preprint arXiv:1710.09412}, 2017.

\bibitem{xu2020automatic}
Ju~Xu, Mengzhang Li, and Zhanxing Zhu,
\newblock ``Automatic data augmentation for 3d medical image segmentation,''
\newblock in {\em International Conference on Medical Image Computing and
  Computer-Assisted Intervention}. Springer, 2020, pp. 378--387.

\bibitem{buslaev2020albumentations}
Alexander Buslaev, Vladimir~I Iglovikov, Eugene Khvedchenya, Alex Parinov,
  Mikhail Druzhinin, and Alexandr~A Kalinin,
\newblock ``Albumentations: fast and flexible image augmentations,''
\newblock {\em Information}, vol. 11, no. 2, pp. 125, 2020.

\bibitem{kavur2021chaos}
A~Emre Kavur, N~Sinem Gezer, Mustafa Bar{\i}{\c{s}}, Sinem Aslan, Pierre-Henri
  Conze, Vladimir Groza, Duc~Duy Pham, Soumick Chatterjee, Philipp Ernst,
  Sava{\c{s}} {\"O}zkan, et~al.,
\newblock ``Chaos challenge-combined (ct-mr) healthy abdominal organ
  segmentation,''
\newblock {\em Medical Image Analysis}, vol. 69, pp. 101950, 2021.

\bibitem{kavur2020comparison}
A~Emre Kavur, Naciye~Sinem Gezer, Mustafa Bar{\i}{\c{s}}, Yusuf {\c{S}}ahin,
  Sava{\c{s}} {\"O}zkan, Bora Baydar, Ula{\c{s}} Y{\"u}ksel, {\c{C}}a{\u{g}}lar
  K{\i}l{\i}k{\c{c}}{\i}er, {\c{S}}ahin Olut, G{\"o}zde~Bozda{\u{g}}{\i} Akar,
  et~al.,
\newblock ``Comparison of semi-automatic and deep learning-based automatic
  methods for liver segmentation in living liver transplant donors,''
\newblock {\em Diagnostic and Interventional Radiology}, vol. 26, no. 1, pp.
  11, 2020.

\bibitem{heimann2009comparison}
Tobias Heimann, Bram Van~Ginneken, Martin~A Styner, Yulia Arzhaeva, Volker
  Aurich, Christian Bauer, Andreas Beck, Christoph Becker, Reinhard Beichel,
  Gy{\"o}rgy Bekes, et~al.,
\newblock ``Comparison and evaluation of methods for liver segmentation from ct
  datasets,''
\newblock {\em IEEE transactions on medical imaging}, vol. 28, no. 8, pp.
  1251--1265, 2009.

\bibitem{antonelli2021medical}
Michela Antonelli, Annika Reinke, Spyridon Bakas, Keyvan Farahani, Bennett~A
  Landman, Geert Litjens, Bjoern Menze, Olaf Ronneberger, Ronald~M Summers,
  Bram van Ginneken, et~al.,
\newblock ``The medical segmentation decathlon,''
\newblock {\em arXiv preprint arXiv:2106.05735}, 2021.

\bibitem{yun2019cutmix}
Sangdoo Yun, Dongyoon Han, Seong~Joon Oh, Sanghyuk Chun, Junsuk Choe, and
  Youngjoon Yoo,
\newblock ``Cutmix: Regularization strategy to train strong classifiers with
  localizable features,''
\newblock in {\em Proceedings of the IEEE/CVF international conference on
  computer vision}, 2019, pp. 6023--6032.

\bibitem{zoph2020learning}
Barret Zoph, Ekin~D Cubuk, Golnaz Ghiasi, Tsung-Yi Lin, Jonathon Shlens, and
  Quoc~V Le,
\newblock ``Learning data augmentation strategies for object detection,''
\newblock in {\em European conference on computer vision}. Springer, 2020, pp.
  566--583.

\bibitem{ghiasi2021simple}
Golnaz Ghiasi, Yin Cui, Aravind Srinivas, Rui Qian, Tsung-Yi Lin, Ekin~D Cubuk,
  Quoc~V Le, and Barret Zoph,
\newblock ``Simple copy-paste is a strong data augmentation method for instance
  segmentation,''
\newblock in {\em Proceedings of the IEEE/CVF Conference on Computer Vision and
  Pattern Recognition}, 2021, pp. 2918--2928.

\bibitem{gong2021keepaugment}
Chengyue Gong, Dilin Wang, Meng Li, Vikas Chandra, and Qiang Liu,
\newblock ``Keepaugment: A simple information-preserving data augmentation
  approach,''
\newblock in {\em Proceedings of the IEEE/CVF Conference on Computer Vision and
  Pattern Recognition}, 2021, pp. 1055--1064.

\bibitem{tian2021self}
Yu~Tian, Fengbei Liu, Guansong Pang, Yuanhong Chen, Yuyuan Liu, Johan Verjans,
  Rajvinder Singh, and Gustavo Carneiro,
\newblock ``Self-supervised multi-class pre-training for unsupervised anomaly
  detection and segmentation in medical images,''
\newblock 2021.

\bibitem{wang2021real}
Xintao Wang, Liangbin Xie, Chao Dong, and Ying Shan,
\newblock ``Real-esrgan: Training real-world blind super-resolution with pure
  synthetic data,''
\newblock in {\em Proceedings of the IEEE/CVF International Conference on
  Computer Vision}, 2021, pp. 1905--1914.

\bibitem{ronneberger2015u}
Olaf Ronneberger, Philipp Fischer, and Thomas Brox,
\newblock ``U-net: Convolutional networks for biomedical image segmentation,''
\newblock in {\em International Conference on Medical image computing and
  computer-assisted intervention}. Springer, 2015, pp. 234--241.

\bibitem{zhou2018unet++}
Zongwei Zhou, Md~Mahfuzur Rahman~Siddiquee, Nima Tajbakhsh, and Jianming Liang,
\newblock ``Unet++: A nested u-net architecture for medical image
  segmentation,''
\newblock in {\em Deep learning in medical image analysis and multimodal
  learning for clinical decision support}, pp. 3--11. Springer, 2018.

\bibitem{sandler2018mobilenetv2}
Mark Sandler, Andrew Howard, Menglong Zhu, Andrey Zhmoginov, and Liang-Chieh
  Chen,
\newblock ``Mobilenetv2: Inverted residuals and linear bottlenecks,''
\newblock in {\em Proceedings of the IEEE conference on computer vision and
  pattern recognition}, 2018, pp. 4510--4520.

\end{thebibliography}

\end{document}